\documentclass[twocolumn,showpacs,preprintnumbers,amsmath,amssymb]{revtex4}
\usepackage{graphicx}
\usepackage{dcolumn}
\usepackage{bm}

\begin{document}

\title{Crossover from mean-field to $2d$ Directed Percolation in the contact process}

\author{T. B. dos Santos$^1$, C. I. N. Sampaio Filho$^1 \footnote{Correspondence to: cesar@fisica.ufc.br}$, Nuno A. M. Ara\'{u}jo$^{1,2}$,  C. L. N. Oliveira$^{1}$, A. A. Moreira$^1$}

\affiliation{$^1$Departamento de F\'{i}sica, Universidade Federal do Cear\'a, 
  60451-970 Fortaleza, Cear\'a, Brasil\\
  $^2$ Departamento  de F\'{i}sica, Faculdade de Ci\^{e}ncias, Universidade de Lisboa, P-1749-016 Lisboa, Portugal, and
  Centro de F\'{i}sica Te\'{o}rica e Computacional, Universidade de Lisboa, P-1749-016 Lisboa, Portugal}

\date{\today}%

\begin{abstract}

We study the contact process on spatially embedded networks, consisting of a regular square lattice with long-range connections. To generate the networks, a long-range connection is randomly added to each node $i$ of a square lattice, following the probability, $P_{ij}\sim{r_{ij}^{-\alpha}}$ , where $r_{ij}$ is the Manhattan distance between nodes $i$ and $j$, and the exponent  $\alpha$ is a tunable parameter. Extensive Monte Carlo simulations and a finite-size scaling analysis for different values of $\alpha$ reveal a crossover from the mean-field to $2d$ Directed Percolation universality class with increasing $\alpha$, in the range $3<\alpha<4$.
\end{abstract}


\maketitle

\section{\label{sec:level1}Introduction} 

The contact process (CP) is the simplest model exhibiting adsorbing-state phase transitions, being considered the starting point for many other methods of nonequilibrium phenomena~\cite{dickmanbook2005}. Spreading of epidemics \cite{satorrasRevModPhys2015,starniniPRL2017},  opinion dynamics \cite{kitsakNatPhys2010}, synchronization \cite{arenas2008}, transport in porous media~\cite{sahimi2011flow}, and phase transitions in driven open quantum spin systems \cite{diehlPRB2017,espigaresPRL2017} are all examples of problems that can be theoretically approached in the CP context. In addition, recent works are showing that even complex experiments, such as, nematic liquid crystals [9] and turbulent flows~\cite{sano2016universal,avilaNATPHYS2016}, can be also explained by contact process models. However, an interesting question that arise is understanding the behavior of the CP phase transition when the contact is no longer bounded by the substract short-range interactions.

Here, we perform Monte Carlo simulations and employ a finite-size scaling analysis to study the critical behavior of the CP on spatially embedded networks with long-range connections~\cite{kleinberg2000,andradePRL2010,havlinNatPhys2011,andradePRE2013}. We consider networks consisting of a two-dimensional square lattice with one additional connection per node~\cite{binderPRE1993,luijtenPRL1996,lubeckPRL2003,lubeckPRE2004,sampaio2013,lucasPRL2014}. This connection is randomly added to connect a node $i$ to a node $j$, according to a probability $P_{ij} \sim{r_{ij}^{-\alpha}}$ that depends on the Manhattan distance $r_{ij}$ between these nodes and on the parameter $\alpha$. The larger the value of $\alpha$ the lower is the probability of having a new connection with a node at a distance $r$. Accordingly, for large values of $\alpha$, the new connections are mainly within nearest and next-nearest neighbors, while for $\alpha=0$ they can be between pair of nodes, regardless of the distance, with equal probability. On these spatially embedded networks, we find a crossover phenomenon, as a function of $\alpha$, from mean-field to $2d$ Directed Percolation (DP) universality class. Furthermore, we show that these regimes are separated by a crossover region where the numerical results are consistent with a continuous change in the value of the critical exponents. A similar rule to construct a spatially embedded networks has been applied to generalize the product rule in percolation process~\cite{sauloPRE2012}. In this case a continuous variation of the critical exponents has been observed that reveals the effect of nonlocality on the scaling properties of the spanning cluster and conducting backbone.

The remaining of the paper is organized as follows. In Section II, we describe the model and the considered  spatially embedded networks. In Section III, the results of our simulations are presented and the finite-size scaling analysis is discussed. Finally, we draw some conclusions in Section IV.
  
\section{\label{sec:level2}The model definition}

In the contact process (CP), each node $i$ of the network is either active $(\sigma_{i} = 1)$ or inactive ($\sigma_{i}=0$). In the absence of diffusion, as considered here, the dynamics evolves through two possible mechanisms: annihilation and creation. In the annihilation, an active node becomes inactive at unit rate, independently of the state of its neighbors. On the other hand, in the creation mechanism, an inactive node that is a neighbor of an active one becomes active at a rate $\lambda/q$, where $q$ is the total number of neighbors of the active node. The state where all nodes are inactive is an adsorbing state and there is a critical creation rate $\lambda = \lambda_{c}$ at which an adsorbing phase transition occurs~\cite{dickmanbook2005}.

We define as the order parameter the density of active nodes, $\rho = \sum_{i}\sigma_{i}/N$, where the sum is over all $N$ nodes. $\rho$ is nonzero for $\lambda> \lambda_{c}$~(active) but vanishes for $\lambda < \lambda_{c}$~(inactive). In purely two-dimensional systems, when long-range connections are absent the critical behavior of the CP is known to fall into the Directed Percolation universality class \cite{dickmanbook2005,lubeckBook2008}. However, above the upper critical dimension ($d>d_{c}=4$), mean-field critical exponents are expected. 

To generate the network, we start with a two-dimensional square lattice, where each node is connected to its four nearest neighbors, and iteratively add to each node
a new connection to one other node of the network, randomly selected according to a probability $P_{ij}{\sim}r_{ij}^{-\alpha}$, where $r_{ij}$ is the Manhattan distance between $i$ and $j$ in the underlying lattice, $\alpha$ is a parameter that controls the length of these long-range connections, and the proportionality factor is a normalization constant that depends on $\alpha$. Our results show that the critical exponents of the CP model on spatially embedded networks depend on $\alpha$. Therefore, the geometric parameter $\alpha$ and the rate of creation $\lambda$ define the two-parameter space for the critical phase diagram of the model.

\begin{figure}[t]
\includegraphics*[width=\columnwidth]{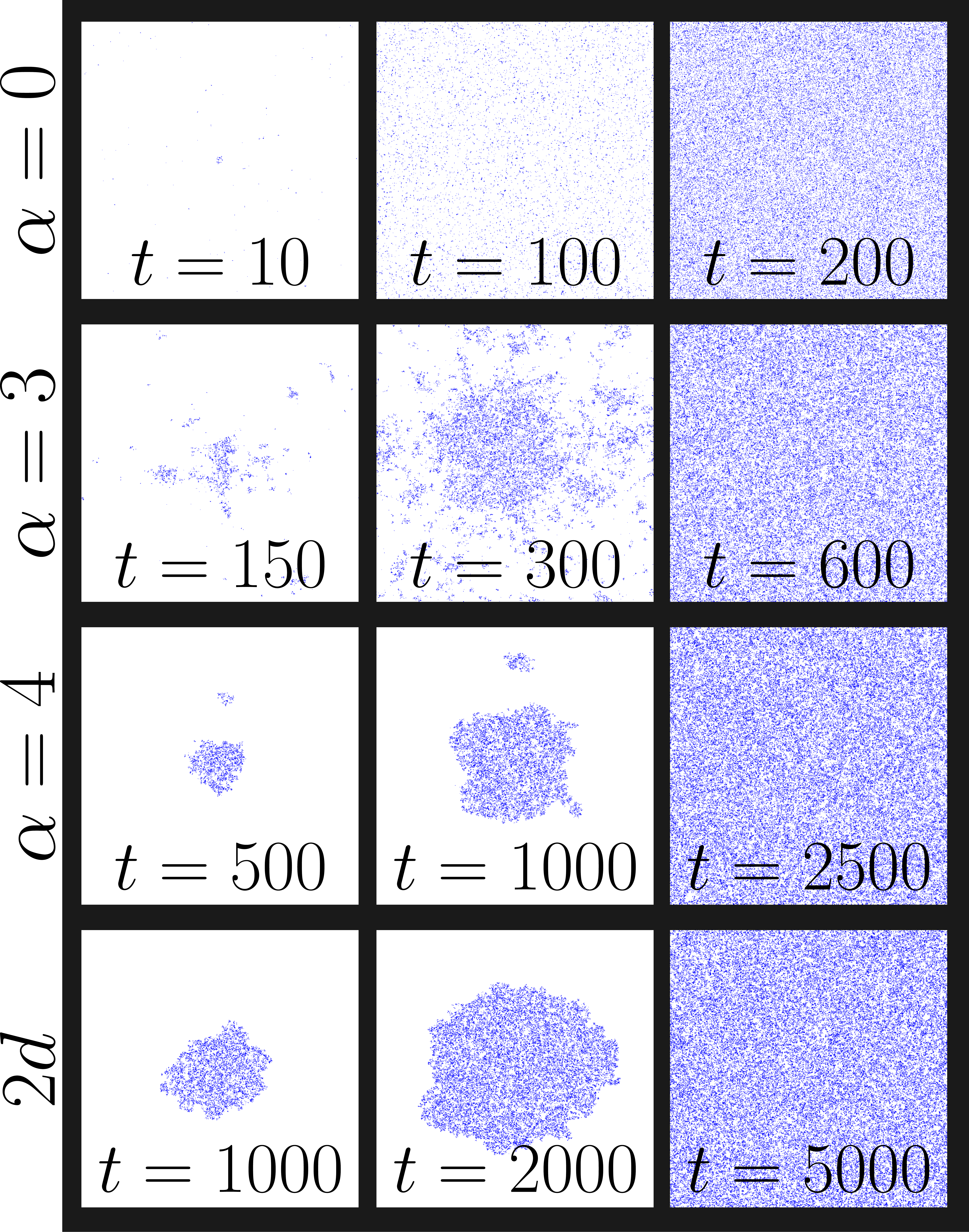}
\caption{Snapshots of a network with $400\times 400$ nodes at different times $t$ and different values of $\alpha$. At the bottom, we show results for purely $2d$ systems, i. e., for regular square lattice with no long-range connections. The initial configuration is the same in all cases: $1\%$ of active nodes, localized at the center of the lattice.}
\label{fig01}
\end{figure}

In order to simulate the CP on spatially embedded networks, we initially define a list of the active nodes and then, iteratively, we select either creation, with probability $\lambda\Delta t$, or annihilation, with probability $\Delta t$, as well as we select an active node at random from the list. In the case of annihilation, the selected node becomes inactive. In the case of creation, a randomly nearest neighbor of the active node is chosen and if it is inactive, then it becomes active. The list of active nodes is then updated accordingly. From the normalization condition, we obtain that $1/\Delta t = 1+\lambda$, and time is incremented by $\Delta t$ after each step. Moreover, we take $\Delta t = 1/N_{p}$, where $N_p$ is the number of active nodes, such that one time unit does not depend on $N_p$~\cite{dickmanPRE2005}.  

The configuration with only inactive nodes is absorbing. Statistically, for a finite system, this configuration is always possible, compromising numerical simulations, especially close to the critical creation rate. To overcome this difficulty, we consider a method proposed in Refs.~\cite{dickmanPRE2005,ferreiraPRE2016}, where one averages only over samples that do not visit an absorbing state. Computationally, we keep a set of $M$ configurations visited previously and, each time the system  gets trapped in the absorbing configuration,
we recover one of the previous $M$ configuration selected at random. Here, we consider $M = 400$ and this catalog of configurations is updated with probability $0.5\Delta t$ at each iteration. 
     
\begin{figure}[t]
\includegraphics*[width=\columnwidth]{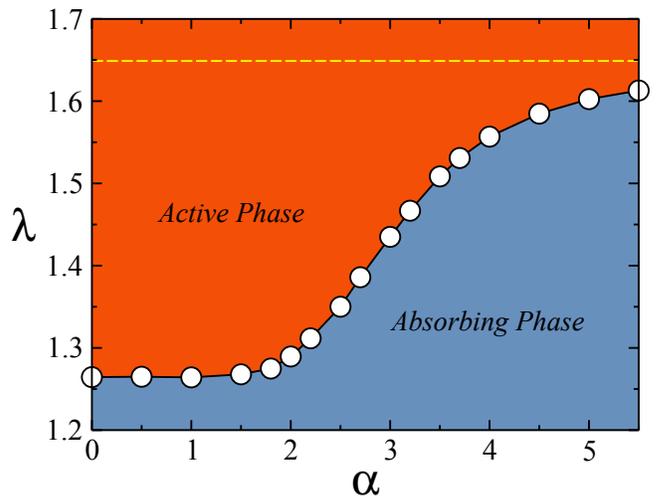}
\caption{Two-parameter phase diagram for the CP on spatially embedded networks. For each value of $\alpha$, the critical creation rate $\lambda_{c}(\alpha)$ (open circles) is obtained from the interception point of the cumulant $\kappa(\lambda)$ (Eq.~\ref{eq00}) for different
system sizes, namely, $N=10000, 22500, 40000, 62500,\mbox{ and } 90000$. $\lambda_{c}$ increases monotonically with the geometric parameter $\alpha$, however, it is always lower the value obtained for a regular square lattice, $\lambda_c = 1.6488\pm 0.0001$ (dashed line)\cite{dickmanbook2005}.}
\label{fig02}
\end{figure}

\section{\label{sec:level3} Results and Discussion}

To characterize the effects of parameters $\lambda$ and $\alpha$ on the phase diagram and on the critical behavior of the CP, we analyze the density of active nodes in the steady state. Figure 1 shows how the concentration of active nodes evolves for different values of $\alpha$. For small values of $\alpha$, it is observed a large spreading of actives nodes and a faster convergence to the steady state, when compared with the regular square lattice, where it is observed a concentric growth of the region of active nodes and a significant slowing down of the dynamics, taking at least one order of magnitude longer in time to reach the stationary state.
  
The phase diagram of the CP is shown in Fig.~\ref{fig02}. For each value of the parameter $\alpha$, the critical value $\lambda_{c}(\alpha)$ is obtained by calculating the ratio~\cite{dickmanPRE1998}
\begin{equation}
\kappa_{N}(\lambda) = \frac{\left\langle \rho^{2}\right\rangle - \left\langle \rho\right\rangle^{2}}{ \left\langle \rho\right\rangle^{2}},
\label{eq00}
\end{equation}
as a function of the creation rate, considering networks with different number of nodes $N$. For sufficiently large networks, we estimate the value of $\lambda_c(\alpha)$ from the value where all curves intercept. As depicted in Fig.~\ref{fig02}, the phase diagram shows that the critical $\lambda_{c}$ increases monotonically with the geometric parameter $\alpha$. For any finite value of $\alpha$, the curve for $\lambda_{c}(\alpha)$ never reaches the value $\lambda_c = 1.6488\pm 0.0001$ corresponding to the critical creation rate on a regular square lattice~\cite{dickmanbook2005}.

\begin{figure}[t]
\includegraphics*[width=\columnwidth]{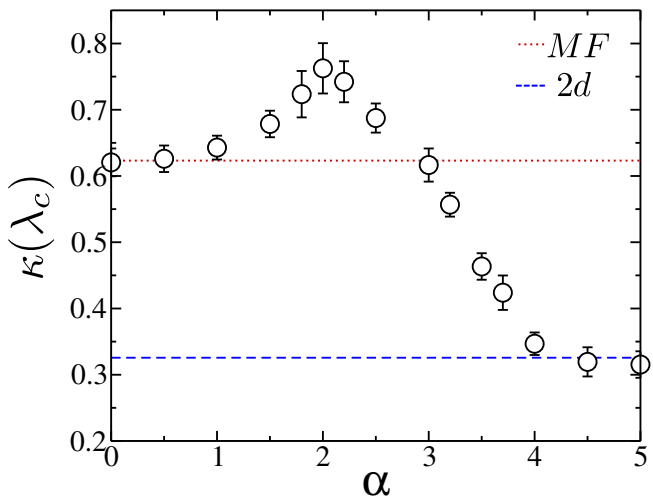}
\caption{Ratio $\kappa(\lambda_{c})$ (Eq.~\ref{eq00}) obtained at the critical creation rate $\lambda_{c}$. The dashed (blue) line corresponding to $0.3275\pm0.0005$ is the value obtained on a regular square lattice~\cite{dickmanPRE1998} while the dotted (red) line corresponding to $0.6232\pm0.0005$ is obtained using the mean-field approach, estimated from simulations of the contact process on a complete graph, with $N = 10000, 20000, 40000, \mbox{ and } 80000$ nodes averaged over $10^{3}$ samples. For some data points the error bars are smaller than the symbols. A maximum of $\kappa(\lambda_c)$ at $\alpha=2$ is reminiscent of the optimal navigation condition with local knowledge~\cite{kleinberg2000}.}
\label{fig03}
\end{figure}

Figure 3 shows the results for the ratio $\kappa(\lambda_{c})$ (open circles) where the curves overlap. For regular square lattices with periodic boundary condition, this ratio is $0.3275\pm0.0005$ \cite{dickmanPRE1998}. In the mean-field regime, here estimated on a complete
graph, we find $\kappa(\lambda_{c}) = 0.6232 \pm 0.0005$ (see caption of Fig.~\ref{fig03} for details). Both the $2d$ and mean-field limits are represented by dashed and dotted lines, respectively, in Fig.~\ref{fig03}. Note that $\kappa(\lambda_{c})$ changes continuously with $\alpha$,  with a maximum when the parameter $\alpha$ equals the dimensionality of the underlying lattice, $\alpha = 2$. Interestingly, our results show that $\kappa(\lambda_c)$ is not limited by the value observed with the mean-field regime, at least for $1<\alpha<3$, where $\kappa(\lambda_c)$ surpasses such regime. In Ref.~\cite{sampaioPRE2016} the authors show also that the Binder cumulant at the critical parameter is $\alpha$-dependent.

\begin{figure}[t]
\includegraphics*[width=\columnwidth]{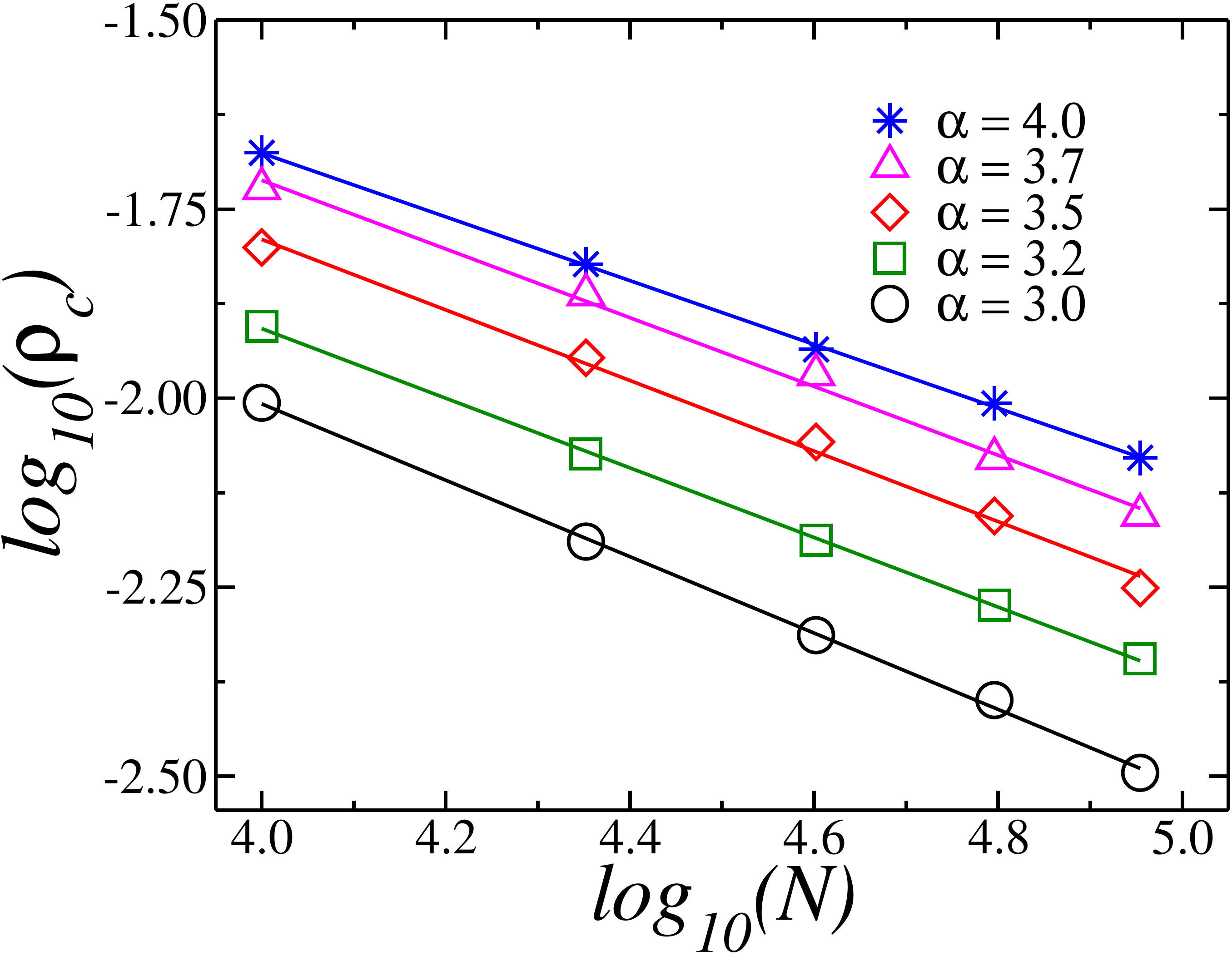}
\caption{Double logarithmic plot showing the size dependence for the order parameter at the critical point, with $\alpha=3.0$ (circles), $3.2$ (squares), $3.5$ (diamonds), $3.7$ (triangles) and $4.0$ (stars). The solid lines represent the least-squares fits to data, whose slopes corresponds to the exponent $\beta/\overline{\nu}$ (see Table \ref{table1}).}
\label{fig04}
\end{figure}

To investigate further the critical behavior of the model, we also perform a finite-size scaling analysis. We assume the following scaling \textit{Ansatz},
\begin{equation}
\rho_{N}(\lambda) \sim N^{-\beta/\overline{\nu}_{\perp}}\widetilde{\rho}(\epsilon N^{1/\overline{\nu}_{\perp}}),
\label{eq03}
\end{equation}
where $\epsilon = (\lambda - \lambda_{c})$ is the distance from the critical creation rate. The exponents $\beta$ and $\overline{\nu}_{\perp}$ are, respectively, associated with the decay of the order parameter $\rho_{N}(\lambda)$ and the divergence of the correlation volume. In the mean-field regime, we have the exact values for these exponents, namely, $\beta = 1$ and $\overline{\nu}_{\perp} = 2$. In the $2d$ regime, however, the exact values are not known but the best numerical estimations are given by $\beta = 0.583 \pm 0.004$ and $\overline{\nu}_{\perp} = 1.466\pm 0.004$ \cite{dickmanbook2005}. Notice that we are defining $\overline{\nu}_{\perp} = d\nu_{\perp}$.

\begin{figure}[t]
\includegraphics*[width=\columnwidth]{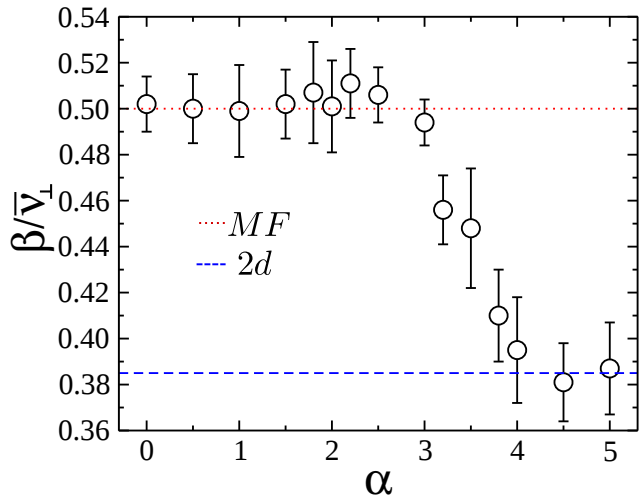}
\caption{The dependence of the critical exponents $\beta/\overline{\nu}$ on the parameter $\alpha$ (see Table \ref{table1}). The upper dotted (red) line corresponds to mean-field exponent, while the bottom dashed (blue) line corresponds to the one for the $2d$ Directed Percolation universality class.}
\label{fig05}
\end{figure}

\begin{table}
\vspace{3mm}
\caption{\label{table1} The estimated values of the critical creation rate $\lambda_{c}$, ratio $\kappa(\lambda_{c})$, and  exponent $\beta/\overline{\nu}$, for the contact process on spatially embedded networks for different values of the parameter $\alpha$. The exponents for the two-dimensional Directed Percolation universality class are $\beta = 0.583\pm 0.004$ and $\overline{\nu}_{\perp} = 1.466\pm 0.004$, while the mean-field exponents are $\beta = 1$ and $\overline{\nu}_{\perp} = 2$.}
\begin{ruledtabular}
\begin{tabular}{ccccccc}
$\alpha$ & & $\lambda_{c}$ & & $\beta/\overline{\nu}_{\perp}$& & $\kappa(\lambda_c)$ \\
0.0 & ~ & $1.2645 \pm 0.0008$ & ~ & $0.502 \pm 0.012$& ~ & $ 0.621 \pm 0.021 $\\
0.5 & ~ & $1.2650 \pm 0.0003$ & ~ & $0.500 \pm 0.015$& ~ & $ 0.626 \pm 0.020 $\\
1.0 & ~ & $1.2642 \pm 0.0006$ & ~ & $0.499 \pm 0.020$& ~ & $ 0.643 \pm 0.018 $\\
1.5 & ~ & $1.2679 \pm 0.0005$ & ~ & $0.502 \pm 0.015$& ~ & $ 0.678 \pm 0.020 $\\
2.0 & ~ & $1.2896 \pm 0.0008$ & ~ & $0.501 \pm 0.020$& ~ & $ 0.763 \pm 0.038 $\\
2.5 & ~ & $1.3501 \pm 0.0006$ & ~ & $0.506 \pm 0.012$& ~ & $ 0.687 \pm 0.022 $\\
3.0 & ~ & $1.4350 \pm 0.0002$ & ~ & $0.494 \pm 0.010$& ~ & $ 0.617 \pm 0.025 $\\
3.2 & ~ & $1.4668 \pm 0.0009$ & ~ & $0.456 \pm 0.015$& ~ & $ 0.557 \pm 0.018 $\\
3.5 & ~ & $1.5085 \pm 0.0003$ & ~ & $0.448 \pm 0.026$& ~ & $ 0.463 \pm 0.020 $\\
3.8 & ~ & $1.5309 \pm 0.0008$ & ~ & $0.410 \pm 0.020$& ~ & $ 0.424 \pm 0.026 $\\
4.0 & ~ & $1.5568 \pm 0.0003$ & ~ & $0.395 \pm 0.023$& ~ & $ 0.347 \pm 0.017 $\\
4.5 & ~ & $1.5848 \pm 0.0012$ & ~ & $0.381 \pm 0.017$& ~ & $ 0.319 \pm 0.022 $\\
5.0 & ~ & $1.6024 \pm 0.0003$ & ~ & $0.387 \pm 0.020$& ~ & $ 0.315 \pm 0.020 $\\
5.5 & ~ & $1.6129 \pm 0.0004$ & ~ & $0.385 \pm 0.021$& ~ & $ 0.312 \pm 0.025 $\\
\end{tabular}
\end{ruledtabular}
\end{table}

Figure 4 is the dependence of the density of active nodes on the network size $N$ at $\lambda = \lambda_{c}$, considering the finite-size scaling relation Eq.(~\ref{eq03}). From this size dependence, we are able to estimate the exponent $\beta/\overline{\nu}_{\perp}$. The results for the critical creation rates and critical exponents, obtained from the simulations for several values of $\alpha$, are summarized in Table~\ref{table1} and Fig.~\ref{fig05}. Within errors bars, we conclude that, for $0 \leq \alpha \leq 3$, the critical exponents are consistent with those of the mean-field critical behavior, whereas for $\alpha > 4$ we recover the exponents of the two-dimensional Directed Percolation (DP) universality class. In the range $3 < \alpha < 4$, we observe an $\alpha$-dependent behavior as a typical signature of a crossover between the mean-field and the DP universality class. These results are consistent with the observed effective dimensionality in spatially embedded networks \cite{sampaioPRE2016,odorSR2015,odorPRE2012,odorPRL2010}. 

\begin{figure}[t]
\includegraphics*[width=\columnwidth]{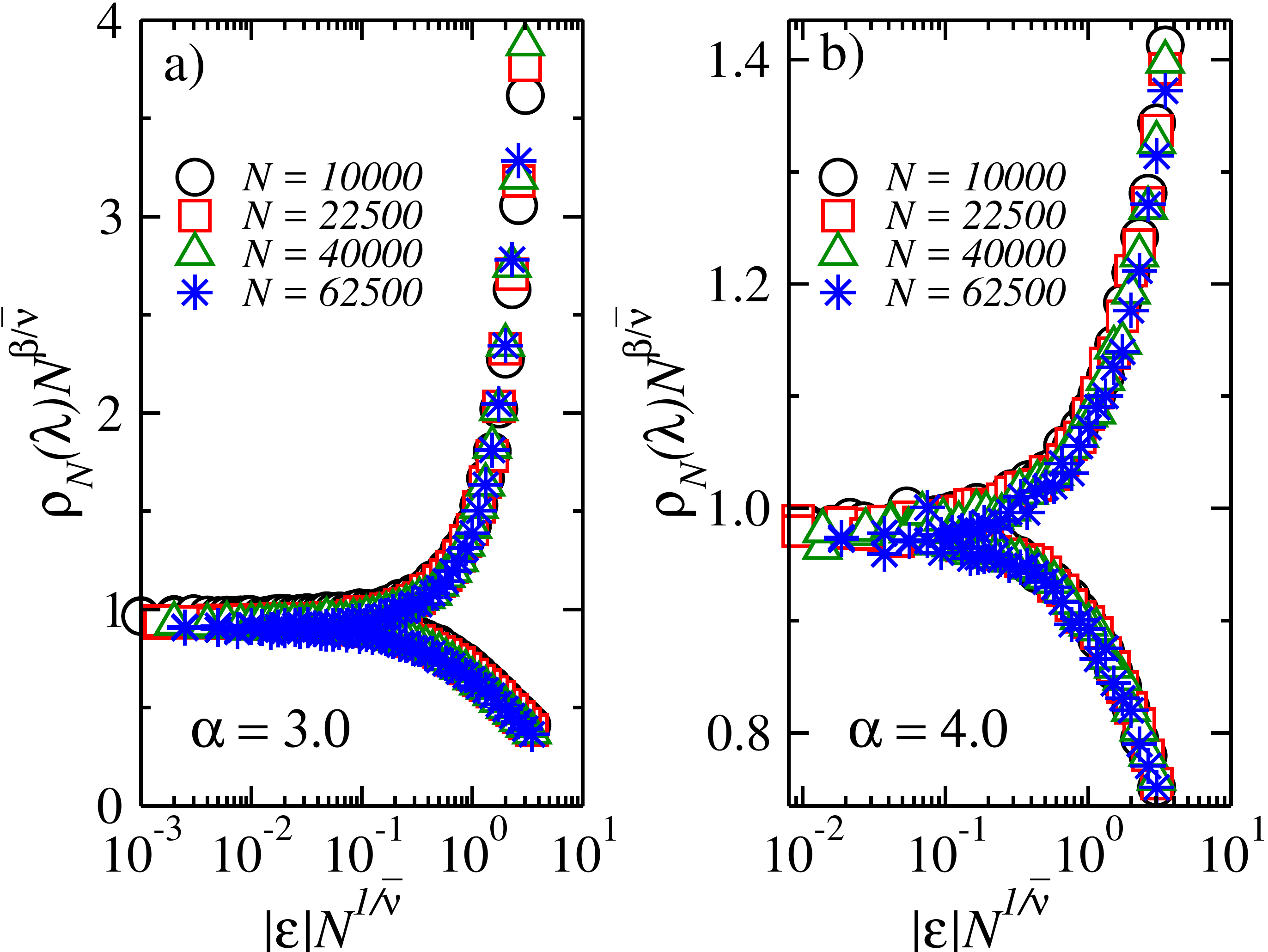}
\caption{Data collapse for the density of active nodes for networks with $N=10000 \mbox{ (circles)}$, $22500 \mbox{ (squares)}$, $40000 \mbox{ (triangles)}$, and $62500 \mbox{ (stars)}$. The universal curve for $\alpha=3$ is consistent with mean-field exponents: $\beta = 1$, $\overline{\nu}_{\perp} = 2$. For $\alpha=4$, the data collapse is obtained using $2d$ Directed Percolation exponents: $\beta = 0.583\pm 0.004$, $\overline{\nu}_{\perp} = 1.466\pm 0.004$.}
\label{fig06}
\end{figure}

To accurately determine the exponents and the nature of the continuous phase transition, we now consider the data collapse of the results from our simulations with different system sizes $N$ for a fixed value of $\alpha$. Figure 6 shows the universal curves for the density of active nodes, where the two regimes can be observed. In Fig.~\ref{fig06}(a) the resulting data collapse is compatible with mean-field critical behavior, consistent with mean-field exponents for $\alpha=3$. Data collapse for $\alpha=4$ is only obtained with two-dimensional DP exponents, as it is shown in Fig.~\ref{fig06}(b).  

\begin{figure}[t]
\includegraphics*[width=\columnwidth]{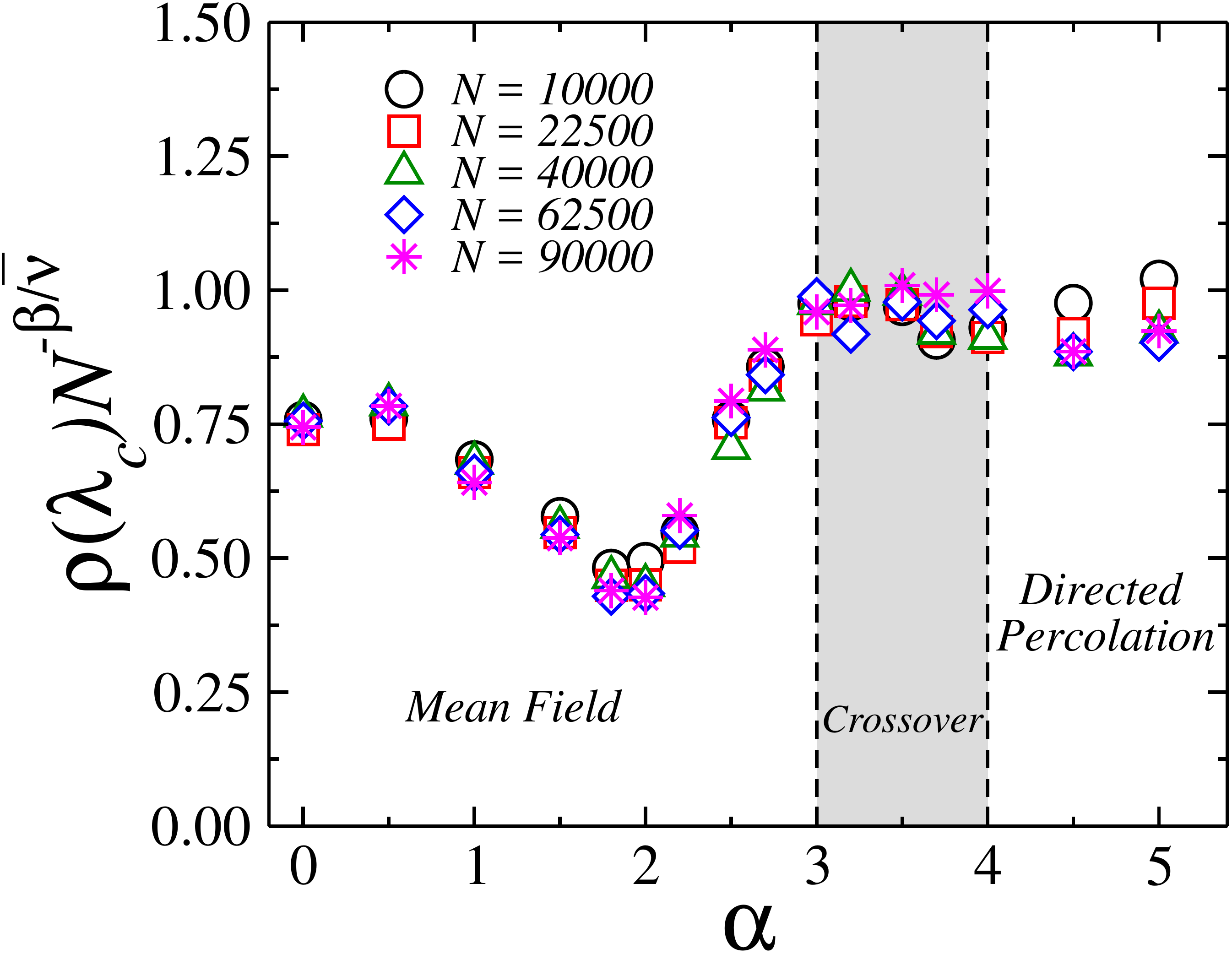}
\caption{Data collapse for the density of actives nodes at the critical creation rate as a function of $\alpha$. The exponents used where: for $\alpha \le 3$, $\beta/\overline{\nu}_{\perp} = 1/2$ (mean-field), for $3<\alpha < 4$ the exponents are $\alpha$-dependent (see Table~\ref{table1}), and for $\alpha\ge 4$,  $\beta/\overline{\nu}_{\perp} = 0.3977\pm0.004$ (Directed Percolation). Notice that a minimum is obtained at $\alpha = 2$.}
\label{fig07}
\end{figure}

Figure~\ref{fig07} shows the data collapse  of the density of active nodes at the critical point as a function of the parameter $\alpha$, considering four values of network sizes $N = 10000, 22500, 40000, 62500$, and $90000$. To collapse the data, we re-scaled the density, $\rho_{N}(\lambda_{c})N^{\beta/\overline{\nu}_{\perp}}$, using the set of calculated exponents, namely, mean-field exponents for $\alpha<3$, $\alpha$-dependent exponents for $3 < \alpha < 4$, and two-dimensional DP exponents for $\alpha>4$. The results shown in Fig.~\ref{fig07} support our conjecture, highlighting the region characterized by non-universal exponents. Moreover, $\alpha = 2$ is associated with the lowest limit for the critical amplitude of the order parameter, differently from the critical amplitude of the ratio, where we observe a maximum.

\section{\label{sec:level4}Conclusions}

In this work, the effects of nonlocal connections on the phase diagram and critical behavior of the contact process (CP) on spatially embedded networks are determined by Monte Carlo simulations and finite-size scaling analysis. The parameters of the model are the creation rate $\lambda$ and the geometric parameter $\alpha$ related to the strength of long-range connections. The resulting phase diagram in the $(\alpha$,$\lambda)$-parameter space indicates that the critical creation rate, $\lambda_c\left(\alpha\right)$, above which the system is in an active state  increases with $\alpha$. The ratio $\kappa_{N}$ calculated at the critical point, whose value has been usually considered as an indicative of a given class of universality, yields results above the mean-field line as $\alpha$ varies in the interval $1 < \alpha <3$,  reaching a maximum value at $\alpha=2$. Nevertheless, for $0 \le \alpha \le 3$ the critical exponents are consistent with those from the mean-field regime. On the other hand, for $\alpha\ge 4$, the calculated values of critical $\kappa(\lambda_{c})$ ratio and critical exponents are both indicative of a network where the dynamics is in the two-dimensional Directed Percolation universality class. Finally, in the region $3 < \alpha < 4$, a continuum crossover can be observed from mean-field to DP critical behavior, which suggests that the CP on spatially embedded networks is described by $\alpha$-dependent exponents.

\begin{acknowledgments}
We thank the Brazilian agencies CNPq, CAPES, FUNCAP, and the National Institute of Science and Technology for Complex Systems for financial support. NA acknowledges financial support from the Portuguese Foundation for Science and Technology (FCT) under Contract no. UID/FIS 00618/2013. 
\end{acknowledgments}

\end{document}